\documentclass[aps,prl,twocolumn]{revtex4}

\usepackage{bm}
\usepackage{amsmath}
\usepackage{graphicx}

\newcommand{\beq}{\begin{equation}}
\newcommand{\eeq}{\end{equation}}
\newcommand{\beqar}{\begin{eqnarray*}}
\newcommand{\eeqar}{\end{eqnarray*}}

\newcommand{\sgn}{\text{sgn}\,}

\newcommand{\ua}{\uparrow}
\newcommand{\da}{\downarrow}

\newcommand{\rtarr}{\rightarrow}

\newcommand{\pb}{\bm{p}}
\newcommand{\rb}{\bm{r}}
\newcommand{\Ab}{\bm{A}}
\newcommand{\Sb}{\bm{S}}
\newcommand{\Rb}{\bm{R}}

\newcommand{\dg}{\dagger}

\newcommand{\lan}{\langle}
\newcommand{\ran}{\rangle}

\newcommand{\Sz}{\langle S_z \rangle}
\newcommand{\Spsq}{\langle S_{\perp}^2 \rangle}
\newcommand{\Szsq}{\langle S_z^2 \rangle}
\newcommand{\Sxsq}{\langle S_x^2 \rangle}
\newcommand{\Sysq}{\langle S_y^2 \rangle}

\newcommand{\TAG}{T_{AG}}
\newcommand{\Tco}{T_{c0}}
\newcommand{\Tc}{T_{c}}

\newcommand{\muB}{\mu_B}

\newcommand{\Gasf}{\Gamma_{\text{sf}}}

\newcommand{\gaorb}{\gamma_{\text{orb}}}
\newcommand{\nuinf}{\nu_\infty}

\newcommand{\nuso}{\nu_{\text{so}}}

\newcommand{\om}{\omega}

\newcommand{\al}{\alpha}
\newcommand{\be}{\beta}

\newcommand{\Ga}{\Gamma}
\newcommand{\de}{\delta}

\newcommand{\la}{\lambda}
\newcommand{\sig}{\sigma}

\newcommand{\eps}{\varepsilon}

\newcommand{\Hc}{{\cal H}}

\newcommand{\lt}{\left}
\newcommand{\rt}{\right}

\begin{document}
\title{Enhancement of Superconductivity
in Disordered Films by Parallel Magnetic Field
}
\author{M. Yu. Kharitonov and  M. V. Feigel'man}
\affiliation{L. D. Landau Institute for Theoretical Physics, Moscow 119334, Russia}
\begin{abstract}
We show that the superconducting transition  temperature $T_c(H)$
of a very thin highly disordered film with strong spin-orbital scattering
can be increased by parallel magnetic field $H$.
This effect is due to  polarization of magnetic impurity spins
which reduces the full exchange scattering rate of electrons;
the largest effect is predicted for spin-$\frac12$ impurities.
Moreover, for some range of magnetic impurity concentrations
the phenomenon of {\it superconductivity  induced by magnetic field}
is predicted: superconducting transition temperature $T_c(H)$
is found to be nonzero in the range of magnetic fields $0 <  H^* \le H \le H_c$.

\end{abstract}
\maketitle
The problem of superconducting alloys with magnetic
impurities was addressed long ago by Abrikosov and
Gor'kov (AG) \cite{AG}. They have shown that
superconductivity (SC) is suppressed
due to exchange scattering (ES) of electrons on magnetic
impurities, the transition temperature $T$ determined from
the equation (hereafter, we employ units, in which $\hbar=1$):
 \beq
  \ln \frac{\Tco}{T}= \pi T \sum_{\eps}
   \lt(\frac{1}{\vert \eps \vert}- \frac{1}{|\eps|+\nu_S} \rt).
 \label{eq:AG}
 \eeq
Here $\eps = 2\pi T(m+1/2)$ is the fermionic Matsubara frequency
($m$ is integer),
$\Tco$ is the transition temperature of clean sample, and
$\nu_S=2\pi N_F n_S J^2 S(S+1)$
is the ES rate of electrons on magnetic
impurities
 ($N_F$ is the normal metal density of states
  per single spin state,
 $n_S$ is the concentration of magnetic impurities,
 $J$ is the exchange coupling constant, and $S$ is
 the impurity spin length).
The solution of~(\ref{eq:AG}) yields the
function $T=\TAG(\nu_S)$.
 \begin{figure}
  \centering
  \includegraphics[width=0.35\textwidth]{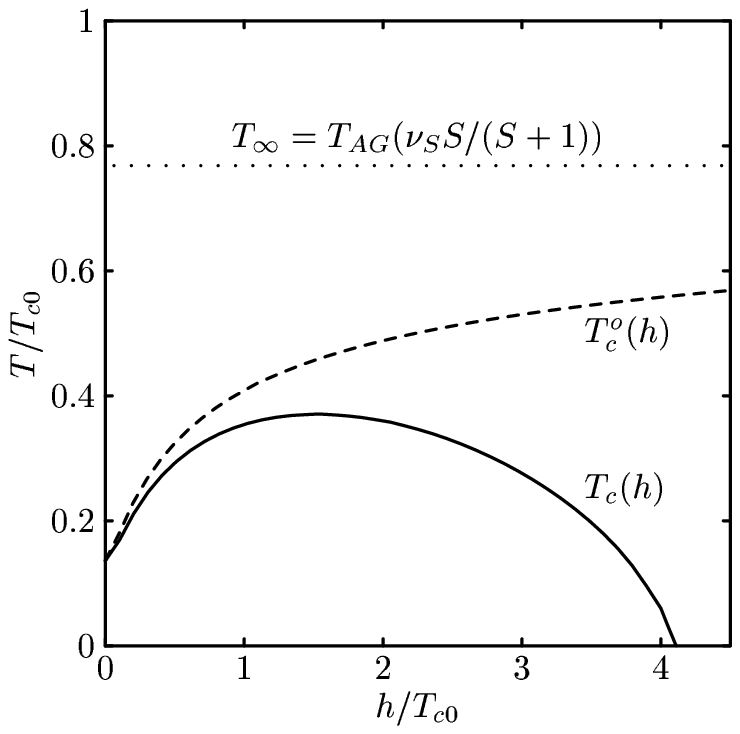}
  \caption{\label{fig:main}
  Enhancement of superconductivity
  by magnetic field.
  $\Tc(h)$ is the transition temperature as a function of
  magnetic field with PE and OE taken into account (solid line).
  The area under $\Tc(h)$ curve corresponds to superconducting state.
  $\Tc^\circ(h)$ is the transition temperature
  with PE and OE disregarded (dashed line),
  $T_\infty = \Tc^\circ(\infty)$ (dotted line).
The parameters used:
$\nu_S=0.85\,\Tco<\nu_S^*$,
$S=1/2$,
$J<0$ (ferromagnetic exchange), $\zeta=5$, $\nuso=10^3\,\Tco$, $\nu=10^4\,\Tco$,
$p_F d=30$.
  $\Tc(0)=0.135\,\Tco$, $T_\infty=0.768\,\Tco$.}
 \end{figure}
 \begin{figure}
  \includegraphics[width=0.35\textwidth]{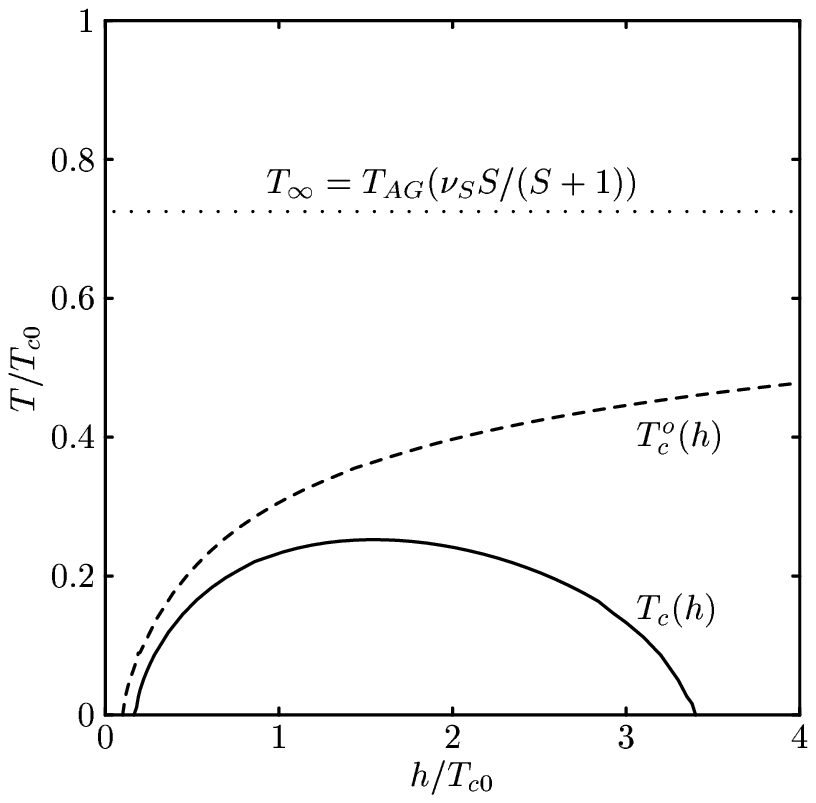}
  \caption{\label{fig:lowT} Magnetic-field-induced
  superconductivity.
  $\Tc(h)$ is the transition temperature as a function of
  magnetic field with PE and OE taken into account (solid line).
 The area under $\Tc(h)$ curve corresponds to superconducting state.
  $\Tc^\circ(h)$ is the transition temperature
  with PE and OE disregarded (dashed line);
  $T_\infty = \Tc^\circ(\infty)$ (dotted line).
The parameters used:
$\nu_S=1.0\,\Tco>\nu_S^*$,
$S=1/2$,
$J<0$ (ferromagnetic exchange), $\zeta=5$, $\nuso=10^3\,\Tco$, $\nu=10^4\,\Tco$,
$p_F d=30$.
   $h^*=0.17\,\Tco$, $T_\infty=0.725\,\Tco$.}
 \end{figure}
There exists a critical point at which the transition
temperature is suppressed down to zero, the critical
scattering rate being
$ \nu_S^*=\pi/(2e^C)\, \Tco=0.882\, \Tco$, where $C=0.577$ is the Euler
constant. The critical concentration, corresponding to $\nu_S^*$,
is further denoted by $n_S^*$.
We emphasize that $\nu_S$ is the {\em full}
ES rate, i.e. the sum of the
spin flip scattering rate $2\pi N_F n_S J^2 (\Sxsq+\Sysq)=2/3\,\nu_S$ and
the rate of scattering without spin flip $2\pi N_F n_S J^2 \Szsq=1/3\,\nu_S$.

The AG's results were derived for unpolarized magnetic impurity spins.
In this Letter we investigate
how the polarization of impurity spins
affects the ES mechanism of
SC suppression.
We show that polarization of
magnetic impurity spins by external magnetic
field reduces the full ES rate $\Ga(\eps)$.
It reaches its minimal value $\nuinf=\nu_S S/(S+1)<\nu_S$
at the infinite field, when the impurity spins are
completely polarized and spin flip processes have frozen out.
This reduction
is due to quantum fluctuations of
impurity spins, thus it
is strongest for $S=1/2$ and vanishes in the limit
$S \gg 1$.

If ES was {\em the only} mechanism of
SC suppression in nonzero magnetic field $h=\muB H$,
the transition temperature $\Tc^\circ(h)$ would always be higher than
$\Tc(h=0)=\TAG(\nu_S)$, determined by AG's result (\ref{eq:AG}).
$\Tc^\circ(h)$ is a growing function,
approaching the  value
$T_\infty=\TAG(\nuinf)$ at very high fields $h\rightarrow\infty$.
The transition temperature increase $\Tc^\circ(h)-\Tc(0)$
comparable to $T_\infty-\Tc(0)$ is attained at the
field range $h\gtrsim \Tc^\circ(h)$. However, apart from ES,
there are  other mechanisms of SC suppression by magnetic field,
namely, paramagnetic effect (PE) and orbital effect (OE).
Thus, to observe an increase $\Tc(h)>\Tc(0)$ of the actual
transition temperature, PE and OE should be
small compared to ES in the field range $h\sim \Tc(h)$.
Strong reduction of PE is achieved in presence of high
spin-orbital scattering rate $\nuso \gg \Tco$ \cite{MakiSO},\cite{Werthamer},\cite{GorRus}.
OE  is suppressed for a  thin-film
(thickness $d$ shorter than the magnetic length $l_H=\sqrt{c/eH}$)
with parallel orientation of external magnetic field \cite{MakiFilm}.

In this Letter we show that the increase in  the transition temperature
can be observed if two quite stringent conditions
on the smallness of PE and OE are met.
First, the spin-orbit scattering rate $\nuso$ must
be sufficiently high:
\beq
 \nuso/\nu_S \gg \zeta^2.
\label{cond:PE}
\eeq
Here  $\zeta=n_S |J| S/ \nu_S = (2\pi N_F |J| (S+1))^{-1} \gg 1$
is the inverse Born parameter for the exchange scattering.
Second, elastic scattering rate $\nu$ and thickness of the film $d$
must satisfy the condition
\beq
 1 \leq  (p_F d)^2 \ll \nu/\Tco \, ,
\label{cond:OE}
\eeq
where $p_F$ is Fermi momentum.

We distinguish between two different regimes depending on
the value of ES rate $\nu_S$.
If $\nu_S<\nu_S^*$, i.e. there exist a finite transition temperature
$\Tc(0)=\TAG(\nu_S)$ at zero field,  then,
provided that the conditions (\ref{cond:PE}), (\ref{cond:OE}) on PE and OE are met,
the increase $\Tc(h)>\Tc(0)$ of the
transition temperature  in some range of $h$ is expected
(see solid line in Fig.~\ref{fig:main}).
The growth of $\Tc(h)$ at $h \lesssim \Tc(h)$ is due to the reduction of
the full ES rate.
At higher fields
PE and OE inevitably prevail,
leading to complete suppression of SC at some
critical field $h_c$.
The most favorable regime for the observation of $T_c(h)$ increase is
when SC is  significantly suppressed at zero field,
i.e. $\nu_S$ is close to (but smaller than) $\nu_S^*$.
In this case large ratio
$(\Tc^{\text{max}}-\Tc(0))/\Tc(0)$ is expected
(see Fig.~\ref{fig:main}).

The most exotic situation occurs
when
$\nu_S> \nu_S^* > \nuinf = \nu_S S/(S+1)$.
Then at $h=0$ superconductivity is totaly suppressed.
Disregarding PE and OE,
one obtains a finite transition temperature $T_\infty=\TAG(\nuinf)$
at very high fields
(indicated by dotted line in Fig.~\ref{fig:lowT}).
If the conditions (\ref{cond:PE}), (\ref{cond:OE}) are satisfied,
superconductivity does not exist below some critical field~$h^*$,
but it appears at higher  fields $h \ge h^*$.
A nonzero transition temperature $\Tc(h)$
(solid line in Fig.~\ref{fig:lowT})
exists in a range of fields starting from
$h^*$ and terminating at some higher critical
field $h_c$, when PE and OE dominate over ES.
Such behavior is possible in the range
of concentrations $n_S^*<n_S<n_S^{**}$,
where $n_S^{**}$ is smaller than $n_S^*(S+1)/S$
and is determined by the parameters involved in PE
and OE.
The better the conditions (\ref{cond:PE}), (\ref{cond:OE}) are
satisfied, the closer is $n_S^{**}$ to $n_S(S+1)/S$.
The most favorable situation for the
experimental observation of ``magnetic-field-induced superconductivity''
is realized when $n_S$ is only slightly larger than $n_S^*$.
In this case $h^*$ is sufficiently small and the curve $\Tc(h)$
produces a quite steep growth at the fields $h$ just above $h^*$
(see Fig.~\ref{fig:lowT}).
Two specific examples of $T_c(h)$ behaviour are presented in
Figs.~\ref{fig:main} and \ref{fig:lowT}
for  $S=1/2$,  for the following set of parameters:
$J<0$ (ferromagnetic exchange), $\zeta=5$, $\nuso=10^3\,\Tco$, $\nu=10^4\,\Tco$,
$p_F d=30$. The similar set of parameters corresponds,
for example, to the $3 \text{nm}$-thick  PtSi film studied in \cite{Baturina},\cite{BaturinaTh}.

Below we briefly outline the method used to derive the
announced results,
details of our calculations will be presented
in a separate publication \cite{largeart}.

The starting point of our problem is the following Hamiltonian:
\[
 \Hc=\Hc_{BCS}+\Hc_{S}+\Hc_{eS}+\Hc_{eU}.
\]
Here,
\beqar
 \Hc_{BCS} &=&\int \lt\{ \psi^{\dg}_\al\lt( \frac{1}{2m}(\pb-e/c \Ab)^2-\eps_F\rt) \psi_\al+
  \rt. \\ &+& \lt. \frac{\la}{2} \psi^\dg_\al \psi^\dg_\be \psi_\be \psi_\al
   - \psi^\dg_\al \,\sig^z_{\al\be} h\, \psi_\be \rt\} d \rb
\eeqar
is the BCS Hamiltonian which includes the orbital and paramagnetic effects of external
magnetic field on conduction electrons;
\[
 \Hc_{eS} = \int \lt\{ \psi^{\dg}_\al \sum_{a} ( u_S \de_{\al\be} + J (\Sb_a, \sig_{\al\be}))
  \de(\rb-\Rb_a)  \psi_\be  \rt\} d \rb
\]
describes the interaction with magnetic impurities and
$
 \Hc_{S}= - \sum_a \om_S S^z_a
$
is the Hamiltonian of impurity spins in external magnetic field
($\om_S=g_S h= 2h$ is their Zeeman splitting).
Finally,
\[
 \Hc_{eU} =  \int \lt\{ \psi^{\dg}_\al(\rb) \sum_{b} v_{\al\be}(\rb-\Rb_b,\rb'-\Rb_b)
   \psi_\be(\rb')  \rt\} d \rb d\rb'
\]
describes  the scattering of electrons on non-magnetic impurities,
which includes both potential and spin-orbit parts. 
Here $v_{\al\be}(\rb,\rb')$ is the Born amplitude in coordinate representation;
since we work in momentum space, we only need its Fourier
tranform
$v_{\al\be}(\pb,\pb')=u_0 \de_{\al\be}
+i\, v_{\text{so}}/p_F^2 ([\pb,\pb'],\sig_{\al\be})$.
Magnetic and non-magnetic impurities are uniformely
distributed over the sample volume with concentrations
$n_S$ and $n$ respectively.

We solve the problem using the
standard diagrammatic technique for BCS theory
and disordered metals \cite{AG},\cite{AGD}
and employing
the following approximations:
i)~$p_F l \gg 1$, where
$l=v_F/\nu$ is the mean free path for potential scattering;
ii)~Born approximation for impurity  scattering;
iii)~``dirty limit'', i.e.  $\nu \gg \nu_{so} \gg T_c$.

The equation for the transition temperature $T$ can be
obtained in the form
\beq
 \ln \frac{\Tco}{T}= \pi T \sum_{\eps}
  \lt(\frac{1}{\vert \eps \vert}- C_0(\eps) \rt),
\label{eq:mainT}
\eeq
where
$
 C_0(\eps)=1/2(C_{\da\da}^{\ua\ua}-C_{\da\ua}^{\ua\da}+
  C_{\ua\ua}^{\da\da}-C_{\ua\da}^{\da\ua})
$ is the singlet Cooperon component.
In the approximation $ p_F l \gg 1$ the Cooperon is
given by an infinite sum of ladder-type diagrams,
each ``ladder step'' containing an impurity line
and the product of two disorder-averaged normal state Green functions.
The expression for the components of such
Green function with electron spin directed along
($\ua$) the external field $h$ and in the opposite direction
($\da$) reads:
\[
 G_{\ua,\da}^{-1}(\eps,\pb)=i\,\eps - \xi \pm h'
  +\frac{i}{2}(\nu+\nuso +\Ga(\eps)) \sgn \eps \pm
   i \tilde{\nu}_S \sgn \,\eps.
\]
Here, $\nu=2 \pi N_F(n_S u_S^2 + n u_0^2)$ is the potential scattering rate,
$\nuso=2\pi N_F n v_{\text{so}}^2 2/3$ is the spin-orbit scattering rate,
$\tilde{\nu}_S= 2\pi N_F n_S u_S J \Sz $
is the interference contribution between potential and exchange scattering
on magnetic impurities (however, this term is
irrelevant and falls out of the final result),
$h'=h-n_S J \Sz$ is the effective magnetic field acting on electron spins comprised
of the external field $h$ and exchange field of polarized impurities $-n_S J \Sz$.
Hereafter $\lan \ldots \ran$ stands for thermodynamic average over
the states of an isolated impurity spin, subjected to external magnetic field $h$:
$
 \lan \hat{A} \ran = 1/Z \sum_{m=-S}^{S} A_{mm}e^{m\om_S/T}
$,
 $Z=\sum_{m=-S}^S e^{m\om_S/T}$.
Thus,
$$
 \Sz=(S+\frac12) \coth\lt[(S+\frac12)\frac{\om_S}{T}\rt]
 -\frac12 \coth\frac{\om_S}{2T} .
$$
Further,
$
 \Ga(\eps)=\nu_z+\Gasf(\eps)
$
is the full ES rate due to exchange interaction of electrons with polarized
magnetic impurities.
It is given by the sum of the rate of scattering without spin flip
$
 \nu_z= \nu_S \frac{\Szsq}{S(S+1)}
$
and the spin flip scattering rate
\beq
 \Gasf(\eps) = \nu_S\frac{\Spsq }{S(S+1)} -\de\Ga(\eps) ,
\label{eq:Gasf}
\eeq
where
\beq
 \de\Ga(\eps)=\nu_S \frac{\Sz}{S(S+1)}
  T\sum_{|\om|>|\eps|} \frac{2 \om_S}{\om^2+\om_S^2}.
\label{DG}
\eeq
Here $\om=2\pi T n$ is the bosonic Matsubara frequency ($n$ is integer) and
$S_\perp^2=S_x^2+S_y^2$.

We now discuss properties of the full exchange
scattering rate  $\Ga(\eps) = \nu_z+ \Gasf(\eps)$ and then
use the knowledge of this function while determining $T_c(h)$.
For  $ |\eps| \gg \om_S $ at any ratio $\om_S/T$ we have
$ \Gasf(\eps) \approx \nu_S\Spsq/S(S+1) $ and $\Ga(\eps) \approx \nu_S$.
At zero field $\Gasf(\eps)=2/3\,\nu_S$, $\nu_z=1/3\,\nu_S$,
and $\Ga(\eps)=\nu_S$ for any $\eps$.
The full ES rate $\Ga(\eps)\approx \nu_S$
for electrons with energies $|\eps| \gg \om_S$ is
not modified by magnetic field, although $\Gasf(\eps)$ and $\nu_z$ do depend on $h$.

Consider the limit of strong polarization $\om_S \gg T$.
In this case one can replace in (\ref{DG}) the sum over $\om$ by the integral
and obtain
\beq
\Gasf(\eps)=\nu_S \frac{1}{S+1}
 \frac{2}{\pi} \arctan\frac{|\eps|}{\om_S} \quad
 {\rm and} \quad \nu_z=\nu_S S/(S+1).
\label{strong}
\eeq
For electron energies $|\eps| \ll \om_S$ less than Zeeman splitting
$ \Gasf(\eps) \approx \nu_S  \frac{1}{S+1}
 \frac{2}{\pi} \frac{|\eps|}{\om_S} \ll \nu_S
$
reflecting the fact that spin flip processes freeze out
for strongly polarized spins.
Hence,
the full ES rate $\Ga(\eps) \approx \nu_z=\nu_S S/(S+1)<\nu_S$
in a wide range of energies $|\eps|\lesssim \om_S$.
At very strong field $\Gasf(\eps) \to 0$ and
$\Ga(\eps) = \nu_\infty= \nu_S S/(S+1)$
for all $\eps$.
Expressing $\Ga(\eps)$ in the form $\Ga(\eps)=\nu_S-\de\Ga(\eps)$
we see that the full ES rate in nonzero field is always less than
$\nu_S$,
with $\de\Ga(\eps,\om_S)$ for a fixed $\eps$ being
a growing function of $\om_S$
with limiting values
$\de\Ga(\eps,0)=0$, $\de\Ga(\eps,\infty)=\nu_S/(S+1)$.

The Cooperon  can be shown to obey 
the following equation
for $C_0(\eps)$
\beq
 \lt(|\eps|+\Ga(\eps) +\frac12 (\hat{L}_0 - \Gasf(\eps)) + \frac{3 h'^2}{2\nuso}+\gaorb \rt)
  C_0=1.
\label{eq:C0}
\eeq
Here
$
 \gaorb=
  \frac{1}{2} D {\lt( \frac{2e}{c} H \rt)}^2 \frac{d^2}{12} =
  \frac{2}{9} (p_F d)^2 \frac{h^2}{\nu}
$
is the dephasing rate corresponding to OE of magnetic field
($D=\frac13 v_F l$ is the diffusion constant) and
the operator $\hat{L}_0$ acts as
\[
 \hat{L}_0 C_0 (\eps) =\nu_S \frac{\Sz}{S(S+1)}
  \,T \sum_{\om} \frac{2\om_S}{\om^2+\om_S^2} C_0 (\eps-\om).
\]

At zero field $h=0$ it is straightforward to check that
$\hat{L}_0 - \Gasf(\eps)=0$ and $\Ga(\eps)=\nu_S$.
Therefore the solution to (\ref{eq:C0})
is $C_0(\eps)=1/(|\eps|+\nu_S)$ and
one recovers the AG's result (\ref{eq:AG}) for transition temperature.

{\it Enhancement of $T_c$ by parallel field}.
We start our analysis from the case $\nu_S<\nu_S^*$, when
a nonzero transition temperature $\Tc(0)=\TAG(\nu_S)$ exists at zero field.
First we study the equation
\beq
 \lt(|\eps|+\nu_S-\de\Ga(\eps) +\frac12 (\hat{L}_0 - \Gasf(\eps)) \rt) C_0=1
\label{eq:C0ES}
\eeq
leaving in (\ref{eq:C0}) the terms related to
ES only and neglecting PE and OE.
In the limit $h \rtarr \infty$ we get: $\hat{L}_0 \rtarr 0$,
$\Gasf(\eps) \rtarr 0$, $\Ga(\eps) \rtarr \nuinf$, and $C_0(\eps)=1/(|\eps|+\nuinf)$.
Thus in the strong-field limit and
in the absense of PE and OE
the transition temperature would be $T_\infty=\TAG(\nuinf)$
(indicated by dotted line in Fig.~\ref{fig:main}),
which is higher than the zero field value $\TAG(\nu_S)$
since $\nuinf<\nu_S$.
For an arbitrary field solving Eqs. (\ref{eq:mainT}),(\ref{eq:C0ES})
together numerically, one obtains
the transition temperature curve $\Tc^\circ(h)$
with PE and OE disregarded
(dashed line in Fig.~\ref{fig:main}).
Formally, the enhancement of transition temperature compared to AG's zero field
result $\TAG(\nu_S)$ is due to the term $-\de\Ga(\eps)$ in (\ref{eq:C0ES})
those effect is always stronger than the (opposite-sign) effect from the
term
 operator $1/2(\hat{L}_0-\Gasf(\eps))$ in the same equation.

We are now in position to derive the conditions  (\ref{cond:PE}) and (\ref{cond:OE})
for the strengths of paramagnetic and orbital effects compatible with observation of
an increase of the actual transition temperature
$\Tc(h)$.
Indeed, the terms in (\ref{eq:C0}) related to PE and OE
must be sufficiently smaller than the terms responsible for ES
in the relevant fields $h\sim\Tco$:
$
  \lt[ h'(h\sim \Tco) \rt]^2/\nuso \ll \nu_S
$
and
$
 \gaorb(h\sim \Tco) \ll \nu_S.
$
Since we are interested in $\nu_S\sim \Tco$ the
latter condition immediately leads to (\ref{cond:OE}).
Due to Born approximation ($\zeta \gg 1$)
for $h\sim\Tco$ and $T \lesssim \Tco$
the exchange field $n_S J \Sz$
dominates over $h$ in the effective field $h'$
and is of the order of its maximal value $n_S J S$.
Therefore, estimating $h'\sim n_S J S$, we obtain (\ref{cond:PE}).
Thus, provided the conditions (\ref{cond:PE}) and (\ref{cond:OE})
are satisfied, one observes an increase in the transition temperature
$\Tc(h)$ (solid line in Fig.~\ref{fig:main}).


{\it Superconductivity induced by magnetic field}.
Now we turn to the case $\nu_S>\nu_S^*>\nu_\infty$ or,
expressed in terms of magnetic impurity concentrations,
$n_S^*<n_S<n_S^* (S+1)/S$.
First we study the Eqs.(\ref{eq:mainT}),(\ref{eq:C0ES}) neglecting
PE and OE.
Since $\nu_S>\nu_S^*$, the SC is totally suppressed at $h=0$,
but at infinite field one obtains a finite
transition temperature $T_\infty=\TAG(\nu_\infty)$
(indicated by dotted line in Fig.~\ref{fig:lowT}),
because  $\nu_\infty<\nu_S^*$.
This leads to the existence of critical field $h_\circ^*$, below which
SC does not exist at any temperature, but
appears in greater fields $h \ge h_\circ^*$.
The field
$h^*_\circ$ is determined from the equation
\beq
 \int_0^\infty d\eps\,(C_0(\eps,h)-1/(\eps+\nu_S^*))=0
\label{hstar}
\eeq
where $C_0(\eps,h)$ is the solution to (\ref{eq:C0ES}) in
zero temperature limit, and depends on only one parameter $\nu_S$.
The transition temperature $\Tc^\circ(h)$
in the absense of PE and OE (dashed line in Fig.~\ref{fig:lowT})
is a growing function of $h$, starting from the zero value
$\Tc^\circ(h^*_\circ)=0$ at $h_\circ^*$ and tending
to $T_\infty$ as $h\rtarr\infty$.
The critical field $h_\circ^*$ as a function of $n_S$ has the following limiting values:
$h_\circ^* \rtarr 0 $ as $n_S \rtarr n_S^* +0$,
$h_\circ^* \rtarr \infty $ as $n_S \rtarr n_S^*(S+1)/S -0$;
and $h_\circ^* \sim \Tco$ when $n_S$ is close neither to
$n_S^*$ nor to $n_S^*(S+1)/S$.

For magnetic impurity concentrations $n_S$  not very close to $n_S^*(S+1)/S$,
the field $h_\circ^* \lesssim \Tco$.
Then, provided
the  conditions (\ref{cond:PE}) and (\ref{cond:OE}) are met,
the described behavior of transition temperature in the fields $h \sim h_\circ^*$
survives under the action of orbital and paramagnetic effects.
PE and OE  slightly change $h_\circ^*$, making the actual critical field
$h^*$ greater than $h_\circ^*$.
The actual transition temperature curve $\Tc(h)$
(solid line in Fig.~\ref{fig:lowT}) is
close to $\Tc^\circ(h)$ at fields $h \sim h^*$ and
deviates sufficiently only at higher fields when
PE and OE dominate over  ES.
We found the critical field $\om_S^*=g_S h^*$ analytically
(with logarithmic accuracy in $\om_S^*/\nu_S^*$)
for the case when $n_S$ is slightly greater than critical $n_S^*$,
i.e. $\de\nu_S=\nu_S-\nu_S^* \ll \nu_S^*$:
\beq
 \frac{\om_S^*}{\nu_S^*} \ln\frac{\nu_S^*}{\om_S^*}=
 \pi(S+1) \lt[ \frac{\de\nu_S}{\nu_S^*}+\frac{3\,(n_S J S)^2}{2\,\nuso \nu_S^*} \rt]
\label{hstar2}
\eeq

If $h_\circ^* \gg \Tco$, i.e.
$n_S$ is close to $n_S^*(S+1)/S$,
accounting for PE and OE,
even with conditions (\ref{cond:PE}) and (\ref{cond:OE}) fulfilled,
destroys SC in such a high field.
Thus, for such $n_S$ SC is totally suppressed at any field.
This yields that
the regime of ``magnetic-field-induced SC'' actually
exists in a more narrow
(than in the absense of PE and OE)
range of concentrations $n_S^*<n_S<n_S^{**}$,
where $n_S^{**}$ is smaller than $n_S^*(S+1)/S$ and is determined
by the values of parameters involved in PE and OE.

In conclusion, we have predicted the mechanism of superconductivity
enchancement in thin films by external parallel magnetic field.
 The effect is due to the polarization of magnetic impurity spins, which reduces
the full rate of electron exchange scattering. In some range
of magnetic impurity concentrations the phenomenon of
{\it magnetic-field-induced superconductivity} is predicted.
The predicted effect is expected to be observable in very thin
disordered superconductive films containing heavy metals
leading to high spin-orbital scattering rate.  We expect that
similar effect may exist in  superconductive-ferromagnet
thin-film bilayers with spontaneous magnetization parallel to the surface.

We are grateful to T.I.Baturina, Ya.V.Fominov, A.I.Larkin and V.V.Lebedev  
for useful discussions.
This research was supported by the RFBR grant 04-02-16348.

\end{document}